\documentclass[aps]{revtex4}
\usepackage{rotate}
\usepackage{epsfig}
\usepackage{psfrag}

\newcommand \be  {\begin{equation}}
\newcommand \bea {\begin{eqnarray} \nonumber }
\newcommand \ee  {\end{equation}}
\newcommand \eea {\end{eqnarray}}

\begin{document}

\title{The endogenous dynamics of markets: price impact and feedback loops}

\author{Jean-Philippe Bouchaud}
\affiliation{Science \& Finance, Capital Fund Management, 6 Bd Haussmann, 75009 Paris France}

\begin{abstract}
We review the evidence that the erratic dynamics of markets is to a large extent of endogenous origin, i.e. determined by the trading activity itself and not due to the rational processing of 
exogenous news. In order to understand why and how prices move, the joint fluctuations of order flow and liquidity -- and the way these impact prices -- become the key ingredients.
Impact is necessary for private information to be reflected in prices, but by the same token, random fluctuations in order flow necessarily contribute to the volatility of markets.
Our thesis is that the latter contribution is in fact dominant, resulting in a decoupling between prices and fundamental values, at least on short to
medium time scales. We argue that markets operate in a regime of vanishing revealed liquidity, but large latent liquidity, which would explain their hyper-sensitivity to fluctuations. More 
precisely, we identify a dangerous feedback loop between bid-ask spread and volatility that may lead to micro-liquidity crises and price jumps. 
We discuss several other unstable feedback loops that should be relevant to account for market crises: imitation,  unwarranted quantitative models, 
pro-cyclical regulation, etc.
\end{abstract}

\maketitle

\tableofcontents 

\section{Introduction}

Why do asset prices move so frequently and why is the volatility so high? Why do prices move at all?
This is obviously a fundamental question in theoretical economics and quantitative finance, that encompasses other, related issues: what
is the information reflected by prices, and to what extent market prices reflect the underlying economic reality? Do we understand the origin of
crises and crashes?

\subsection{Efficient markets}

The neo-classical paradigm answers that question as follows: prices change because new information about the fundamental value of the asset becomes available. If the
information is instantly and perfectly digested by markets, then prices should reflect faithfully these fundamental values and only
move because of exogenous unpredictable news. This is the Efficient Market story, which assumes that informed rational agents would arbitrage away 
any error or small mispricing, and nudge the price back to its ``true'' value. This is very much a Platonian view of the world where markets merely reveal 
fundamental values {\it without influencing them} -- the volatility is an unbiased measure of the flow of news, and is not related to the trading 
activity {\it per se}. Crashes, in particular, can only be {\it exogenous}, but not induced by market dynamics itself. 

Is this picture fundamentally correct to explain why prices move and to account for the observed value of the volatility? Judging from the literature, 
it looks as if a majority of academics still believe that this story is at least a reasonable starting point. The
idea of rational agents and efficient markets has shaped the mind-set of decision makers and regulators for decades and has permeated a variety of spheres, from international monetary policy to derivative markets or sociology. Scores of financial mathematics papers are deeply rooted in the idea that option markets are efficient. It is standard practice in banks to calibrate unwarranted models using market prices
of so-called liquid derivatives, and use these models to price and hedge other (more exotic) derivative instruments, a practice very much prone to non-linear amplification of errors and self-fulfilling feedback loops. In the aftermath of the crisis, a number of scholars and pundits have expressed concern about this whole intellectual construct, in particular about the intrinsic stability of 
markets (see, among the most provocative ones, \cite{Soros, AnimalSpirits, FG, Marsili}, and in the context of financial markets \cite{Taleb, Wilmott}) -- bearing in mind that Keynes had anticipated a lot of these `new' ideas \cite{Keynes}. 
Alan Greenspan himself appears to have been fooled by the Efficient Market whim. As he recently admitted:  
{\it those of us who have relied on the self-interest of lending institutions, myself included, 
are in a state of shocked disbelief...Yes, I've found a flaw [in the theory]. I don't know how significant or permanent it is. But I've been very distressed by that fact.}. Paul Krugman tried to explain ``how Economists got it so wrong''\cite{Krugman} as follows: {\it As I see it, the economics profession went astray because economists, as a group, 
mistook beauty, clad in impressive-looking mathematics, for truth.}. But in many quarters it is still business as usual -- for
example David Altig, from the Atlanta Fed, declared in September 09: {\it I'm less convinced that we require a major paradigm shift.
Despite suggestions to the contrary, I've yet to see the evidence that progress requires moving beyond the intellectual boundaries in which 
most economists already live.} As we write, it is still common practice in the world of quantitative finance and in the derivative industry to use of blatantly irrelevant models 
(such as the local volatility fallacy \cite{sabr} \footnote{This by the way, probably explains why Dupire's seminal paper on local volatility models \cite{Dupire} 
is among the most cited papers in mathematical finance. I do find this fact particularly symptomatic of the diseases of financial engineering.}, 
the use of Gaussian or Archimedean copulas \cite{copulas}, etc.) that can always be brute force calibrated on market data to 
spit out meaningless numbers. In my view, financial engineering is at the stage of Ptolemea's epicycles before Kepler's ellipses. 
After so much twisting and tweaking (calibration is the politically correct word for it), epicycles were more precise that ellipses...but of course, this was no theory.

There are many reasons to believe that markets are very far from efficient in the above traditional sense. To start with, 
the very concept of a ``fundamental value'', that can be computed, at least as a matter of principle, with arbitrary 
accuracy with all information known at time $t$, 
appears to be deeply flawed. The number of factors affecting the fundamental value of a company (or of a currency, etc.) is so large, and the influence of unknown-unknowns so predominant, that there should be, at the very least, an irreducible error margin. All valuation models or predictive tools used by traders and market participants (using economic ratios, earning forecasts, etc.) or based on statistical analysis that detect trends or mean-reversion, are extremely noisy (statistical methods can only rely on a rather short history) and often even biased. For example, financial experts are known to be on the whole over-optimistic, and rather imprecise at forecasting the next earning of a company (see e.g. \cite{Guedj05} and references therein). 
News are often ambiguous and not easy to interpret, and real information can be buried underneath Terabytes of irrelevant data. 

If we accept the idea of an intrinsically noisy fundamental value with some band within which the price can almost freely wander
(because nobody can know better), the immediate question is: how large is this irreducible uncertainty? Is it very small, say $10^{-4}$ in relative terms, 
or quite a bit larger, say $50\%$ -- as suggested by Black, who defined an efficient market as a market giving the correct price to within a factor 2 \cite{Black86}? If Black is right (which we tend to believe) and the uncertainty in the fundamental value is large, then Keynes' famous 
beauty contest is a better narrative of what is going on in financial 
markets, at least in the short term. It is less the exogenous dynamics (news driven) of the fundamental value than 
the endogenous dynamics of supply and demand that should be the main focus of research. 

Another reason why markets cannot be efficient is the limited intelligence of us humans (even if, quite strangely, many academics have a hard time 
coming to terms with this \footnote{A lot could be said -- and some has been said -- about the religious roots and the political implications of the rational agent concept.}). We do make mistakes and have regrets, and we do make suboptimal decisions. In fact, even perfectly rational agents 
that have to process information in a finite amount of time, are likely to make errors or go for suboptimal solutions. A good illustration of this is provided by chess: pressed by time, even chess masters do make errors and lose against Deep Blue. Many optimisation problems are indeed very complex, in the sense that the best algorithm to solve them requires a time that grows exponentially as a function of 
the size of the problem (for example the size of a portfolio that one wishes to optimise, see \cite{Gall}). Humans just cannot be expected to be any good at such tasks without 
developing intuitive or heuristic rules -- the most common one being: {\it Just do what your neighbour is doing, he might know better}. Another one is: {\it Look for patterns, they might 
repeat} (on this one, see \cite{Arthur,WB}).

\subsection{Market impact}

This in fact leads us to a crucial issue, that of market impact, which is the main theme of the present paper. It is both rather intuitive and
empirically demonstrated that buy trades are followed by a rise of the price, and sell trades by a price decline. A simple way to try to guess what others are doing is to observe price variations, that may reflect the impact of their trades, and therefore their intentions \cite{Kyle}. 
The interpretation of the price impact phenomenon is 
however potentially controversial. In the Efficient Market picture, impact is nearly tautological, since informed agents successfully forecast short term 
price movements and trade to remove arbitrage opportunities. This trivially results in correlations between trades and price changes, but these 
correlations cannot be exploited by copy-cats. In this story, however, uninformed trades should have no price impact (except maybe on short time scales); 
otherwise silly trades would, in the long run, drive prices arbitrary far from fundamental values.

A more plausible story is the following: if Black's idea is correct and the uncertainty in the fundamental value is large, then the amount 
of information contained in any given trade is necessarily small. \footnote{Empirically, the standard deviation of market impact is found to be 
very large compared to its mean, confirming that the quantity of information per trade must indeed be small.} Furthermore, modern electronic markets are anonymous, which makes it impossible to distinguish potentially informed trades from non informed trades. Hence, all trades are equivalent and they must (statistically) equally impact prices. 

The mechanism by which the market reacts to trade by shifting the price is precisely the above copy-cat heuristic rule, applied at a tick 
by tick level. Since all agents are pretty much in the dark but believe (or fear) that some trades might contain useful information, 
prices must statistically move in the direction of the trades. As reviewed below, 
high frequency data allows one to make much more precise statements about the amplitude and time-dependence of this impact. 
But the consequence of such a scenario is that even silly trades do impact prices and contribute to volatility -- a mechanism for instabilities, 
bubbles and crashes, even without any `news' or other fundamental cause for such events. 

We therefore have to decide between two opposite pictures for the dynamics of price: {\it exogenous}, news driven, or {\it endogenous}, impact driven. 
Of course, reality should lie somewhere in the middle. In the next sections, we will review several empirical findings that suggest that endogenous dynamics is in fact dominant in financial markets.

\section{Exogenous or endogenous dynamics?}

Are news the main determinant of volatility? Were this true, and in the absence of ``noise traders'', the price should essentially be constant
between news, and move suddenly around the release time of the news. Noise traders should merely add high frequency,
mean-reverting price changes between news, that do not contribute to the long term volatility of the price. 

There are, however, various pieces of evidence suggesting that this picture is fundamentally incorrect. First, high frequency time series do not
look at all like long plateaus dressed by high frequency noise. On liquid assets, there is very little sign of high frequency mean reversion that one could attribute to noise traders -- in other words, 
the high frequency volatility is very close to its long term asymptotic value (see e.g. \cite{subtles}). \footnote{Here we talk about the 
volatility of the mid-point, not of the traded price, that shows a large, trivial bid-ask bounce.}
Volatility is furthermore well known to be much too high to be explained by changes in fundamentals \cite{Shiller}, and most large price swings seem to
be unrelated to relevant news release. This was the conclusion reached by Cutler, Poterba and Summers in a seminal study of large daily price changes \cite{Poterba} (see also \cite{Fair} for a more recent discussion with identical conclusions). 

\subsection{News and no-news jumps}

We have recently confirmed in detail this conclusion, now on high frequency data, 
using different news feeds synchronised with price time series. 
We have looked for simultaneous occurrences of price ``jumps'' and intra-day news releases on a given company \cite{Joulin}.  \footnote{Overnight news and
overnight jumps are not included in the study. `Big' company news are usually issued overnight. But this makes the existence of intraday jumps all the
more puzzling!} This requires one
to define jumps in a consistent, albeit slightly arbitrary fashion. We chose to compare the absolute size $|r(t)|$ 
of a one minute bin return to a short term (120 minutes) flat moving average of the same quantity,  $\sigma(t)$, in order to factor in local modulations
of the average volatility. An s-jump is then defined such that $|r(t)| > s \, \sigma(t)$. The number of s-jumps as a function of $s$
is shown in Fig. 1; it is seen to decay as $\approx s^{-4}$, consistent with previous work on the distribution of high frequency returns \cite{Gopi,Gabaix}. 
We note once again that this distribution is very broad, meaning that the number of extreme events is in fact quite large. 
For example, for the already rather high
value $s=4$ we find 7 to 8 jumps per stock per day! 
A threshold of $s=8$ decreases this number by a factor $\approx 10$, amounting to one jump every one day and 
a half per stock. On the same period, we find on average one news item every 3 days for each stock.
These numbers already suggest that a very large proportion of shocks cannot be attributed to idiosyncratic
news (i.e. a news item containing the ticker of a given stock). This conclusion still holds when one includes 
(possibly also endogenous) collective market or sector jumps in the definition of news. The number of jumps explained by these
`macro' events only increases by $20 \%$, but leaves most jumps unexplained (see \cite{Joulin} for more details). One may also
argue that these jumps are due to the arrival of private information. But this cannot be since an investor really possessing superior
information will {\it avoid} disturbing the market by trading too quickly, in order not to give away his advantage. As illustrated by 
Kyle's model \cite{Kyle}, an insider should better trade incrementally and discretely. We will discuss below strong empirical evidence that 
trading indeed occurs incrementally.

\begin{figure}[htbp]
\centering
\resizebox{10cm}{!}{\includegraphics[trim = 0 -40 0 -60]{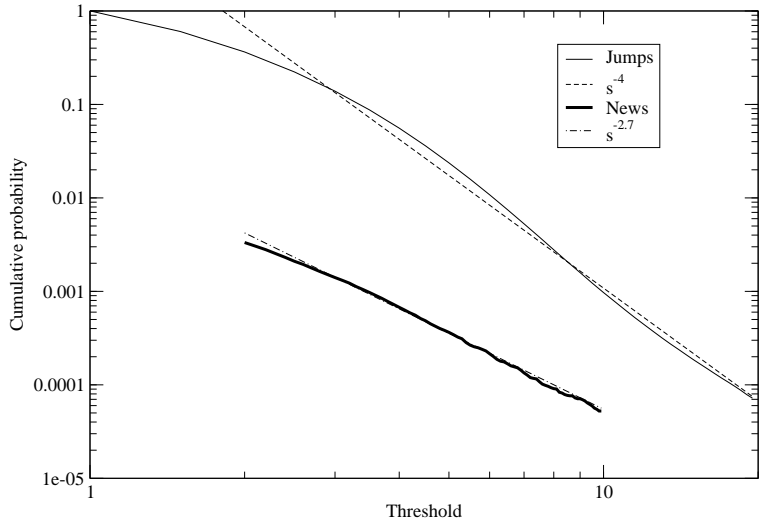}}
\caption{Cumulative distribution of s-jumps as a function of $s$. The scale is log-log. The distribution decays as $\approx s^{-4}$. We also show the 
number of s-jumps associated with news. Interestingly, this distribution also decays as a power-law but with a smaller exponent, 
$\approx s^{-2.7}$. From \cite{Joulin}}
\label{jump_distrib}
\end{figure}

More quantitatively, there are striking statistical differences between jumps induced by news, and jumps with no news, that clearly demonstrate that the 
two types of events result from genuinely distinct mechanisms. One difference resides in the distribution of jump sizes: as shown in Fig. 1, the cumulative 
distribution of jumps with news has again a power law tail $s^{-\mu}$, 
but with an exponent $\mu \approx 2.7$, different from the value $\mu = 4$ mentioned above for jumps without news. 
Interestingly, if we extrapolate these
distributions deep in the tail (and far beyond the observable regime), the news induced jumps eventually become more probable than the no-news jumps, but 
only for $s \approx 60$!

\begin{figure}[htbp]
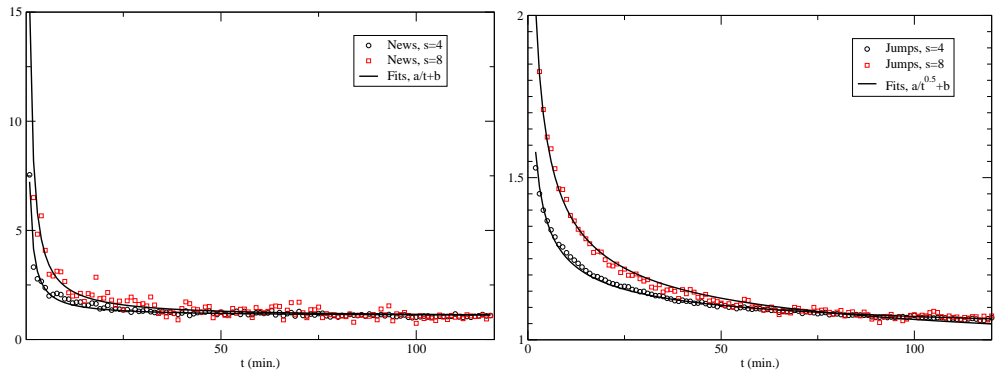

    \begin{tabular}{cc}  
    \includegraphics[width=6.5cm]{Newsrelax.eps}&
    \includegraphics[width=6.5cm]{Jumprelax.eps}
    \end{tabular}
    \caption{{\bf a)} Relaxation of the volatility after $s=4$ and $s=8$ news jumps, and power law fit with an exponent $\zeta=1$.
    {\bf b)} Relaxation of the volatility after $s=4$ and $s=8$ jumps, and power law fit with an exponent $\zeta=1/2$. From \cite{Joulin}.}
   \label{fig:vol2}
\end{figure}

A second difference is the way the volatility relaxes after a jump. In both cases, we find (Figure~\ref{fig:vol2}) that the relaxation of the excess-volatility follows a power-law
in time $\sigma(t)-\sigma(\infty) \propto t^{-\zeta}$ (as also reported in \cite{Kertecz,Mantegna}). The exponent of the decay is, however, markedly different in the two cases: for news 
jumps, we find $\zeta \approx 1$, whereas for no-news jumps one has $\zeta \approx 1/2$, with in both cases little dependence on the value of the
threshold $s$. The difference between endogenous and exogenous volatility relaxation has also
been noted in \cite{Muzy}, but on a very restricted set of news events. Although counter-intuitive at first, 
the volatility after a no-news jump relaxes more slowly 
than after a news. This could be due to the fact that a jump without any clear explanation makes traders anxious for a longer time than if a well identified event caused the jump. The slow, non-exponential relaxation after a no-news jump is very interesting per se, and already suggests 
that the market is in some sense critical. 

So, yes, some news do make prices jump, sometimes a lot, but the jump frequency is much larger than news frequency, meaning that most intraday 
jumps appear to be endogenous, induced by the speculative dynamics itself that may  spontaneously cause liquidity micro-crises. 
In fact, a decomposition of the volatility (made more precise in Sect. \ref{IV-A} below) into
an impact component and a news component, confirms this conclusion: most of the volatility seems to arise from trading itself, 
through the very impact of trades on prices.  

\subsection{Universally intermittent dynamics}

Another striking observation, that could be naturally accounted for if price movements do result from the endogenous dynamics of a complex system, is the {\it universality} of many empirical stylised facts, such as the Pareto tail of the distribution of returns, or the intermittent, long memory nature 
of the volatility. These features are observed across the board, on all traded, liquid assets, and are quantitatively very similar. 
We show for example in Fig. 3 the distribution of the relative daily changes of the 60-day implied volatility corresponding to the S\&P100 stocks 
from 1st Jan. 2001 to 1st Jan. 2006 \cite{Biely}. There is a priori no reason whatsoever to expect that the statistics of implied volatility returns should resemble that of price returns. The implied volatility represents the market consensus on 
the expected volatility of the stocks for the 60 days to come. But as Fig. 3 illustrates, the distribution of implied volatility returns has the very 
same shape as that of any
other traded asset, whatever its nature. In particular, the positive and negative tails of the distribution decay here as $|r|^{-4}$ -- 
very much as the tails of the daily price returns of stocks. The Pareto exponent is always found to be in the same ballpark for any liquid asset
(stocks, currencies, commodities, volatilities, etc.). This suggests again that
these tails are not generated by strong exogenous shocks, but rather by the trading activity itself, 
more or less independently of the nature of the traded asset.

\begin{figure}[htbp]
\centering
\resizebox{10cm}{!}{\includegraphics[trim = 0 -40 0 -60]{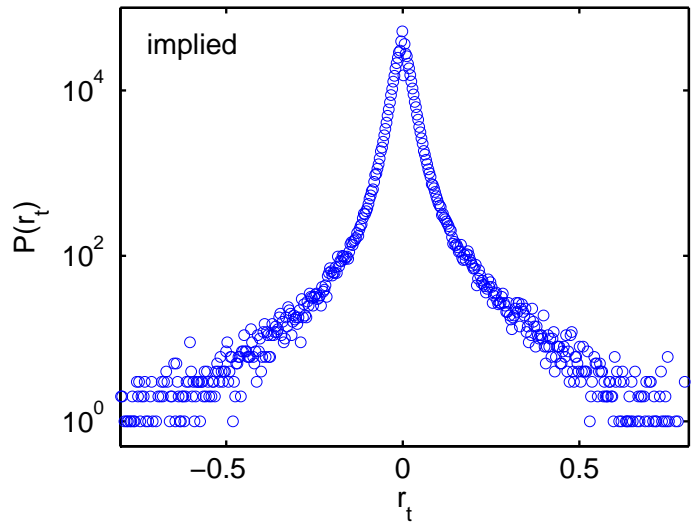}}
\caption{Semi-log plot of the probability distribution of the daily returns $r$ of the 60-day implied volatility corresponding to the S\&P100 stocks 
from 1st Jan. 2001 to 1st Jan. 2006. The positive and negative tails of the distribution decay  as $|r|^{-4}$, much as the distribution of the underlying 
stock returns.}
\label{implied}
\end{figure}

The activity and volatility of markets have a power-law correlation in time, reflecting their intermittent nature (see Fig. 3): quiescent 
periods are intertwined with bursts of activity, on all time scales. Interestingly, many ``complex'' physical systems display very similar intermittent dynamics \cite{PW}: 
velocity fluctuations in turbulent flows \cite{Frisch}, avalanche dynamics in random magnets under a slowly varying external field \cite{Sethna}, teetering progression 
of cracks in a slowly 
strained disordered material, etc. \cite{PLD} (see also \cite{others} for related papers). 
The crucial point about all these examples is that while the exogenous driving force is regular and steady, the resulting endogenous dynamics is complex and jittery. These systems find a temporary equilibrium where activity is low before reaching a tipping point
where avalanches develop, leading to a new quasi-equilibrium -- sometimes close to the previous one, sometimes very far. In financial
markets, the flow of ``real'' news is of course needed to stir the activity, but the scenario we favour is similar: it is the {\it response} of the
market that creates turbulence, and not necessarily the cause, barring of course exceptional 
events that do sometimes severely disrupt markets (for example, Lehman's
bankruptcy). As explained above, these events are however much too rare to explain why prices jump so frequently.  

\begin{figure}[htbp]
\centering
\resizebox{10cm}{!}{\includegraphics[trim = 0 -40 0 -60,angle=270]{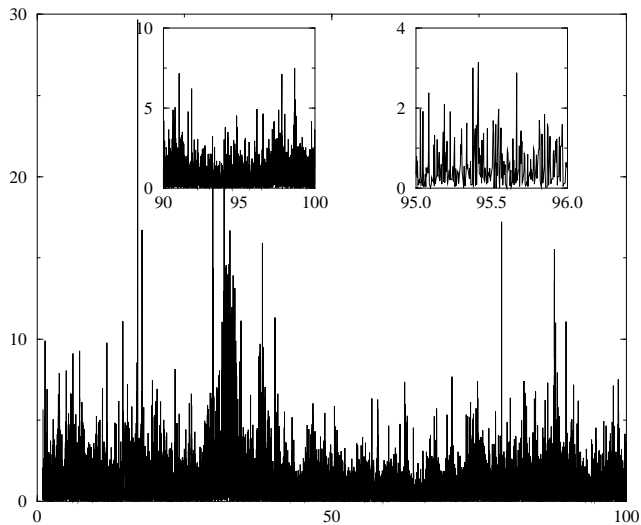}}
\caption{Absolute value of the daily price returns for the Dow-Jones index over a century (1900-2000), and zoom on different
scales (1990-2000 and 1995). Note that the volatility can remain high for a few years (like in the early 1930's) or for 
a few days. This volatility clustering can also be observed on high frequency (intra-day) data.}
\label{intermittence}
\end{figure}

In all the above physical examples, the non-trivial nature of the dynamics comes from collective effects: individual components have a relatively simple behaviour, 
but interactions lead to new, emergent phenomena. Since 
this intermittent behaviour appears to be generic for physical systems with both heterogeneities and interaction,
it is tempting to think that the dynamics of financial markets, and more generally of economic systems, do reflect the same underlying mechanisms. 
We will come back to these ideas in the conclusion.

\section{Are markets in ``equilibrium?''}

Recent access to UHF, tick by tick data allows one to investigate the microscopics of order flow and price formation. As we will explain 
below, the analysis of these data sets calls for a substantial revision of the traditional view of the Walrasian {\it t\^atonnement} process,
that in theory should allow prices to quickly settle to their equilibrium values 

\subsection{Trades are long ranged correlated!}

Each transaction can be given a sign $\epsilon=\pm 1$ according to whether the trade took place at the ask, and was triggered by a buy
market order, or at the bid, corresponding to a sell market order. Market orders cross the half-spread and are usually interpreted as resulting
from agents possessing superior information that urge them to trade rapidly, at the expense of less informed traders who place limit orders. Whether 
or not this interpretation is correct, it is an empirical fact that such market orders impact prices, in the sense that there is some clear correlation
between the sign of a trade and the following price change. The impact function is a quantitative measure of this, and is defined as:
\be\label{Rdef}
{\cal R}(\ell)=\langle (p_{n+\ell}-p_n) \cdot \epsilon_n \rangle_n,
\ee
where $p_n$ is the mid-point price immediately preceding the $n$th trade, and the average is taken over all trades, independently of their volume.

The efficient market story posits that each trade is motivated by a new piece of information, which quickly moves the price towards its 
new value. Since by definition the direction of the news is unpredictable, the resulting string of signs $\epsilon_n$ should have very short 
range correlations in time. The surprising empirical result discovered by \cite{subtles,Farmer04} (see \cite{BFL} for a review) is that 
the autocorrelation of the sign of trades is in fact very {\it{long-range correlated}}, over several days or maybe even months. The 
sign correlation function decays extremely slowly, as a power-law: 
\be
{\cal C}(\ell) \equiv \langle \epsilon_{n+\ell} \cdot \epsilon_n \rangle_n \propto \ell^{-\gamma},
\ee
where the exponent $\gamma$ is found to be around $0.5$ for stocks and around $0.8$ for futures. The fact that these binary strings have long-memory 
(in the sense that $\gamma < 1$) turns out to have 
important technical consequences, discussed below. The long-memory nature of the sign process means that the order flow is highly predictable. 
Conditional on observing a buy trade now, one can predict with a rate of success a few percent above $\frac12$ that the sign of the $10,000$th
trade from now (corresponding to a few days of trading) will be again positive!

\subsection{Scant liquidity and trade fragmentation}

Where does such a persistent correlation come from? A crucial point is that even ``highly liquid" markets are in fact not that liquid. 
Take for example a US large cap stock. Trading is extremely frequent: tens of thousands of trades per day, adding up to a daily volume of roughly $0.1 \%$ of total market capitalisation. However, the volume of buy or sell limit orders typically available in the order book {\it at a given instant of time} is quite small: only of the order of $1 \%$ of the traded daily volume, i.e. $10^{-5}$ of the market cap for stocks.  
Of course, this number has an intra-day pattern and fluctuates in time, and it can reach much smaller values during liquidity crises. 

The fact that the outstanding liquidity is so small has an immediate consequence: trades must be fragmented. It is not uncommon that investment funds want to buy large fractions of a company, often several percents. If trading occurs through the continuous double auction market, the numbers above suggest  that  to buy $1 \%$ of a company requires at least the order of $1000$ individual trades.  It is clear that these trades have to be diluted over several days, since otherwise the market would be completely destabilised, leading to unacceptable costs for an aggressive buyer.  Thus if an investment fund has some information about the future price of a stock, it  cannot use it immediately, and has to trade into the market incrementally in order to avoid paying its own impact \cite{Kyle}. This fragmentation of orders clearly leads to long-range correlations in the sign of trades (see \cite{BFL,Wspread} for a more thorough discussion of the empirical evidence for this fragmentation interpretation, rather than a copy-cat mechanism, at least on long time scales). Trade fragmentation is a direct evidence that most investors are, to some degree, insensitive to price changes. Once the decision to buy has been made, the trade is completed even if the price moves up and down, at least within some bounds on the order of a few days or a few weeks of volatility. This is in line 
with the idea that the inherent uncertainty on the price is rather large. 

\subsection{Markets slowly digest new information}

From a conceptual point of view, the most important conclusion of this qualitative discussion is that prices are typically not in equilibrium, in the traditional Marshall sense.  That is, the true price is very different than it would be if it was such that supply and demand were equal -- as measured by the honest intent of the participants, as opposed to what they actually expose.  As emphasised above, because of `stealth trading', the volume of individual trades is much smaller than the total demand or supply at the origin of the trades. This means that most of the putative information is necessarily latent, withheld by participants because of the small liquidity of the market. Information can only slowly be incorporated into prices (see \cite{Lyons01} for similar ideas). Markets are hide and seek games between ``icebergs'' of unobservable buyers and sellers, that have to 
funnel through a very small liquidity drain. Prices cannot be instantaneously in equilibrium. At best, the notion of equilibrium prices can only make sense when coarse-grained over a long time scale, 
but then the flow of news, and the change of prices themselves, may alter the intention of buyers and sellers.  

But why is liquidity, as measured by the number of standing limit orders, so meager? Because ``informed'' traders that would use limit orders are reluctant 
to place large orders that would reveal their information. Liquidity providers who eke out a profit from the spread are also reluctant to place large limit orders that put them at risk of being `picked-off' by an informed trader.  Buyers and sellers face a paradoxical situation:  both want to have their trading done as quickly as possible, but both try not to show their hands and reveal their intentions.  As a result, markets operate in a regime of vanishing {\it revealed liquidity}, but large {\it latent liquidity}. 

The long-range nature of the sign correlation however leads to a beautiful paradox. As we emphasised above the sign of the order flow is highly 
predictable. Each trade furthermore impacts the price in the direction of the trade. How come, then, prices can remain statistically efficient in the
sense that there is hardly any predictability in the sign of price changes? The resolution of this paradox requires a more detailed description of
the impact of each trade, and in particular the time dependence of this impact. This is what we address in the next section.

\section{Impact and resilience}

We qualitatively discussed the origin of price impact in the introduction section. Even at this microlevel, one is faced with the exogenous vs. endogenous debate about the origin of price changes. In the efficient market picture, 
as emphasised by Hasbrouck \cite{Hasbrouck}, ``orders do not {\it impact} prices. It is more accurate to say that orders {\it forecast} prices.'' However, if the market collectively believes that even a small fraction of trades 
contain true information, the price will on average be revised upwards after a buy and downwards after a sell. But while impact is a necessary mechanism 
for information to be reflected by prices, its very existence means that ``information revelation'' could merely be a self-fulfilling prophecy, which 
would occur even if the fraction of informed trades is in fact zero. 

\subsection{Some empirical facts about impact}
\label{IV-A}
In any case, using high frequency data, one can measure impact accurately. The average change of mid-point between two successive transactions,
conditioned to a buy trade or after a sell trade are found to be equal to within error bars:
\be
E[ p_{n+1}-p_n | \epsilon_n=+1] \approx - E[ p_{n+1}-p_n | \epsilon_n=-1] = {\cal R}(\ell=1),
\ee
where we have used Eq. (\ref{Rdef}) for the definition of the {\it instantaneous impact}. Note that in the definition above we average over 
all trades, independently of their volume. It is well known that the dependence of impact on volume is very weak: it is more the trade itself, rather than its 
volume, that affects the price \cite{Jones,BFL}.  This is often interpreted in terms of discretionary trading: large market orders are only submitted when there 
is a large prevailing volume at the best quote, a conditioning that mitigates the impact of these large orders. 

One important empirical result is that the impact ${\cal R}(\ell=1)$ is proportional to the bid-ask spread $S$: ${\cal R}(\ell=1) \approx 0.3 S$.
This proportionality holds both for a given stock over time, as the spread fluctuates, and across an ensemble of stocks with 
different average spreads. This law means that the market instantaneously updates the valuation of the asset almost to the last traded price 
(in which case one would find ${\cal R}(\ell=1) \approx \frac{S}{2}$). 

What happens on longer time scales? A plot of ${\cal R}(\ell)$ vs. $\ell$ reveals that
the impact first grows with time by a factor two or so in the first $100-1000$ trades, before saturating or maybe even reverting back \cite{subtles}. 
However, the interpretation of this
increase is not immediate since we know that the signs of trades are correlated: many trades in the same direction as the first trade will occur.
From this point of view, it is even surprising that ${\cal R}(\ell)$ does not grow more than a factor of two. This is related to the paradox
mentioned above.  

Another remarkable empirical finding is that the volatility per trade $\sigma_1$ is found to be proportional to the instantaneous impact, and
therefore to the spread. In fact, if one regresses the volatility per trade as a function of the impact, as:
\be
\sigma_1^2 = A {\cal R}_1^2 + {\cal J}^2,
\ee
where ${\cal R}_1={\cal R}(\ell=1)$ and ${\cal J}^2$ is the contribution of news induced jumps, that should happen with very little trading, one finds that this latter contribution is
very small compared to the first, see \cite{Wspread}. The relation between $\sigma_1$ and $S$  is again true both for a single stock over time and 
across different stocks. A very simplified picture accounting for this finding is that the spread defines a `grid' over which the price moves with a random direction at every trade. Of course, the problem with this interpretation is that the long-ranged nature of the sign correlations should lead to super-diffusion, i.e. persistent trends in the price  -- we are back to the same paradox. 

\subsection{A subtle dynamical equilibrium}

Let us assume that the price at trade time $t$ can be decomposed as a sum over past impacts, in the following way:
\be
p_t = p_{-\infty} + \sum_{t'=-\infty}^{t-1} G(t-t') \, \epsilon_{t'} \, S_{t'}\,  V_{t'}^\psi,  
\ee
where $S_t$ is the spread at time $t$ and $V_t$ the volume of the trade at that instant. The exponent $\psi$ is found to be quite 
small (see \cite{BFL} for a detailed discussion): as noted above, it is well documented that the response to the volume of a single trade is strongly concave. The most important quantity
in the above equation is the function $G(\ell)$, that ``propagates'' the impact of the trade executed at time $t'$ up to time $t$. 
In other words, $G$ can be
interpreted as the impact of a single trade, in contrast to ${\cal R}$ that sums up the impact of correlated trades. Within the above model, the
relation between the two quantities reads:
\be
{\cal R}(\ell) =  K \left[G(\ell)+ 
\sum_{0 < n < \ell} G(\ell-n) {\cal C}(n) + 
\sum_{n > 0} \left[G(\ell+n)-G(n)\right] {\cal C}(n)\right],
\ee
where $K$ is a certain constant, and ${\cal C}$ is the correlation of the signs of the trades (see \cite{subtles}). If impact was permanent, i.e. 
$G(\ell)=G_0$, the long range nature of the correlation of trades would lead to an ever growing ${\cal R}(\ell)$, as $\ell^{1-\gamma}$ for $\gamma < 1$, 
i.e. for a long memory process. Whenever $\gamma > 1$, ${\cal R}(\ell \gg 1)$ saturates to a constant. This underlies the significance of the 
fact that empirically $\gamma$ is found to be less than unity.

If on the other hand $G(\ell)$ decays as $\ell^{-\beta}$ with $\beta$ exactly tuned to $(1-\gamma)/2$, then the transient nature of the impact 
of single trades precisely offsets the long range correlation of the sign of trades. This choice of $\beta$ leads to 
both a saturating ${\cal R}(\ell)$ and a diffusive
price, for which returns are uncorrelated \cite{subtles}. The solution of our paradox is therefore that the market is resilient: after the immediate reaction to 
a trade, the impact slowly mean-reverts back to zero (but in the mean time, of course, new trades occur).

The above model can be  reformulated in terms of surprise in order flow. Since the order flow is highly correlated, the past history of 
trades allows one to make a prediction of the sign of the next trade, that we call $\widehat \epsilon_t$. Within a linear filter framework, 
this prediction can be expressed in terms of past realized signs:
\be
\widehat \epsilon_t = \sum_{t'=-\infty}^{t-1} B(t-t') \, \epsilon_{t'}, 
\ee
where $B(\ell)$ are coefficients. If we forget the fluctuations of the product $SV^\psi$  \footnote{It would not be difficult to 
include them in a model where the whole product $\epsilon SV^\psi$ follows a similar linear regression model on its past values.},
it is easy to show that the above transient impact model can be exactly rewritten in terms of a permanent response to the surprise in the
order flow, defined as $\epsilon_t - \widehat \epsilon_t$:
\be\label{surprise}
p_t = p_{-\infty} + G(1) \sum_{t'=-\infty}^{t-1} \left[\epsilon_{t'}-\widehat \epsilon_t\right],
\ee 
provided the following identification is made: $G(1)B(\ell)=G(\ell+1)-G(\ell)$ (see \cite{Gerig07,BFL}). If $B(\ell)$ corresponds to the best 
linear filter adapted to the long-ranged correlation in the $\epsilon$s, one easily recovers that $G(\ell)$ indeed 
decays as $\ell^{-\beta}$ with $\beta=(1-\gamma)/2$.

The above interpretation in terms of surprise is interesting because it provides a microscopic mechanism for the decay of the impact of single trades. 
Let us rephrase the above result in more intuitive terms. Call $p_+ > \frac12$ the probability that a buy follows a buy. 
The unconditional impact of a buy is $G(1)$ (see Eq. \ref{surprise}). From the same equation, a second buy immediately following the first has a reduced impact, $G^+(1) < G(1)$, since now $\widehat \epsilon = 2p_+ - 1 > 0$. 
A sell immediately following a buy, on the other hand, has an enhanced impact equal to 
$G^-(1) > G(1)$. If we want the next trade to lead to an unpredictable price change, one must have that 
the its conditional average impact is zero: $p_+ G^+(1) - (1-p_+) G^-(1) \equiv 0$, which indeed leads to $G^-(1) = \frac{p_+}{1-p_+} G^+(1) > G^+(1)$
when $p_+ > \frac12$ \cite{Gerig07}. 
This is the ``asymmetric liquidity" effect explained in \cite{Farmer04,Farmer06,Gerig07}. This mechanism is expected to be present in general: 
because of the positive correlation in order flow, the impact of a buy following a buy should be less than the impact of a sell following a sell -- otherwise trends would build up. 

Now, what are the mechanisms responsible for this asymmetric liquidity, and how can they fail (in which case markets cease to be efficient, and jumps appear)? 
One scenario is ``stimulated refill'': buy market orders trigger an opposing flow of sell limit orders, 
and vice-versa (\cite{subtles}). This rising wall of limit orders decreases the probability of further upward moves of the price, 
which is equivalent to saying that $G^+(1) < G^-(1)$. This dynamical feedback between market orders and limit orders is therefore fundamental for the stability of 
markets and for enforcing efficiency. It can be tested directly on empirical data; for example the authors of \cite{Weber05} have found strong evidence 
for an increased limit order flow compensating market orders, see also \cite{Eisler} for similar results. 

This stabilisation mechanism can be thought of as a dynamical version of the supply-demand equilibrium, in the following sense: 
incipient up trends quickly dwindle because as the ask moves up, market order buy pressure goes down while the limit order sell pressure increases (see also \cite{Mad}). 
Conversely, liquidity induced mean-reversion -- that keeps the price low -- attracts more buyers, which is in turn an incentive for liquidity providers to raise their price. 
Such a balance between liquidity taking and liquidity providing is at the origin of the subtle compensation 
between correlation and impact explained above. In fact, the relation  between volatility and spread noted above is a direct manifestation of the
very same competition between market orders and limit orders \cite{Wspread}. Limit orders are only profitable if the spread is larger than the volatility, whereas 
market orders are profitable in the opposite case. A small spread attracts market orders, whereas a large spread attracts limit orders. In orderly
market conditions, an equilibrium is reached, enforcing $\sigma_1 = c S$, where $c$ is a numerical constant \cite{Wspread}. But this tight relation
can also lead to an instability: a local increase of volatility leads to an opening of the spread, itself feeding back on volatility. This
mechanism might be at the heart of the frequent liquidity micro-crises observed in markets, and the associated no-news jumps reported above.
The relation between volatility and spread means that there is a kind of `soft-mode': the market can operate at any value of the volatility, provided 
the spread is adapted (and vice-versa). The absence of a restoring force pinning the volatility to a well defined value is probably responsible for 
the observed long-memory property, and the slow relaxation of the volatility after a jump (see Fig. 2).

\subsection{The problem with impact}

In conclusion, although ``price impact'' seems to convey the idea of a forceful and intuitive mechanism, the story behind it might not be that simple. Empirical studies show that
the correlation between signed order flow and price changes is indeed strong, but the impact of trades is neither linear in volume nor permanent, as assumed
in several models. Impact is rather found to be strongly concave in volume and transient (or history dependent), the latter property being a necessary consequence
of the long-memory nature of the order flow. 

Coming back to Hasbrouck's comment, do trades {\it impact} prices or do they {\it forecast} future price changes? Since trading on modern electronic markets is anonymous,  
there cannot be any obvious difference between ``informed'' trades and ``uninformed'' trades. 
Hence, the impact of any trade must statistically be the same, whether informed or not informed. 
Impact is necessary for private information to be reflected in prices, but by the same token, random fluctuations in order flow must necessarily contribute to the volatility of markets.
As argued all along this paper, our belief is that the latter contribution is significant, if not dominant. 

\section{Summary and Perspectives}

Let us reiterate the main points of the present paper, which aimed at describing why and how asset prices move and identifying the the building blocks of any quantitative model that claims to reproduce the dynamics of markets.

\subsection{Markets are close to a critical point}

We have first made a strong case that the dynamics of markets is mostly endogenous, and determined by the trading activity itself. 
The arguments for this are: 
\begin{itemize}
\item a) news play a minor role in market volatility; most jumps appear to be unrelated to
news, but seem to appear spontaneously as a result of the market activity itself; 
\item b) the stylised facts of price statistics (fat-tails in the distribution of returns, 
long-memory of the volatility) are to a large extent universal, independent of the particular nature of the traded asset, and very reminiscent of 
endogenous noise in other complex systems (turbulence, Barkhausen noise, earthquakes, fractures, etc.). In all these examples, the intermittent, avalanche nature of the dynamics is an emergent property, unrelated to the exogenous drive which is slow and regular. 
\end{itemize}

In search of a purely endogenous interpretation of these effects, it is natural to investigate to high-frequency, micro-structure ingredients that 
generate price changes. 
We have discussed the remarkable long-range correlations in order flow that has far-reaching consequences and forces us to revise many 
preconceived ideas about equilibrium. First of all, these correlations reflect the fact that even ``liquid'' markets are in fact very illiquid, in the
sense that the total volume in the order book available for an immediate transaction is extremely small ($10^{-5}$ of the market capitalisation for stocks). 
The immediate consequence is that the trades of medium to large institutions can only be executed incrementally, explaining the observed correlation in the order flow. 
By the same token, the information motivating these trades (if any) cannot be instantaneously reflected by prices. Prices cannot be in equilibrium, 
but randomly evolve as the icebergs of latent supply and demand progressively reveal themselves (and possibly evolve with time). This feature is an 
unavoidable consequence of the fact that sellers and buyers must hide their intentions, while liquidity providers only post small volumes in fear of 
adverse selection.

The observation that markets operate in a regime of vanishing revealed liquidity, but large latent liquidity is crucial to understand their 
hyper-sensitivity to fluctuations, potentially leading to instabilities. Liquidity is necessarily a dynamical phenomenon that reacts to order flow such as to dampen the trending effects and keep price returns unpredictable, through the subtle `tug-of-war' equilibrium mentioned above. 
Such a dynamical equilibrium can however easily break down. For example, an upward fluctuation in buy order flow might trigger a momentary panic, 
with the opposing side failing to respond immediately. Similarly, the strong structural link between spread and volatility can ignite a positive feedback loop whereby increased spreads generate increased volatility, which itself causes liquidity providers to cancel their orders and widen the spread.
Natural fluctuations in the order flow therefore lead, in some cases, to a momentary lapse of liquidity, explaining the frequent occurrence of price jumps without news. An extreme incarnation of this feedback loop probably took place during the ``flash crash'' of May 6th, 2010. 
We believe that the formal limit of zero liquidity is a critical point \cite{Bak}, which would naturally explain the analogy between the dynamics of markets and that of other complex systems, in particular the universal tails and the intermittent bursts of activity.  We are however lacking a precise model that would allow one to formalise these ideas (see \cite{MG,Mike} for work in that direction).

In summary, the picture of markets we advocate is such that the lion's share of high frequency dynamics is due to fluctuations in order flow. 
News and information about fundamental values only play the role of ``stirring'' the system, i.e. slowly changing the large latent supply and demand, 
except on relatively rare occasions where these events do indeed lead to violent shocks. Most of the market activity comes from the slow
execution of these large latent orders, that cascades into high frequency fluctuations under the action of the use of liquidity providers 
and liquidity takers, who compete to exploit all statistical regularities. 

The end product of this activity is a white noise signal. Prices are, in a first approximation, statistically efficient in the sense that there is little predictability left in the time series. But this does not necessarily mean that 
these prices reflect in any way some true underlying information about assets. We believe, as Keynes and Black did, that the uncertainty in fundamental values is so large that there
is no force to anchor the price against random high frequency perturbations. It is quite remarkable indeed 
that the high frequency value of the volatility 
approximately coincides with the volatility on the scale of weeks, showing that there is very little mean-reverting effects to rein the high frequency tremor of markets.
Only when prices reach values that are -- say -- a factor 2 away from their ``fundamental value'' will mean-reverting effects progressively come into play. 
In the context of stocks, this only happens on the scale of months to years, see \cite{Thaler} and the discussion in \cite{WB}. From this point of view, as emphasised by Lyons \cite{Lyons01}, 
``micro-structure implications may be long-lived'' and ``are relevant to macroeconomics''.

\subsection{Looking forward}

Having said all this, the theoretical situation is still rather disappointing. There is at this stage no convincing framework to account for these effects, in the sense 
of converting the above qualitative ideas into a quantitative model that would, for example, predict the shape of the tails of the return distribution, or the long-range
memory of volatility after a suitable coarse-graining in time. This is in my mind the most interesting research program in quantitative finance: build models from bottom up such that the value and the dynamics of the parameters (volatility, correlations, etc.) can be estimated, or at least understood. Most available models to date 
(agent based models \cite{Lux,collective}, Minority Games \cite{MG}, herding models \cite{Cont}, Langevin approaches \cite{Lang}, etc.)  postulate a linear (in volume) and permanent impact as
in the Kyle model \cite{Kyle}, 
whereas as we have shown, impact is both non-linear and
transient. It may well be that the assumption of a linear, permanent impact is justified after some coarse-graining in time, say on a daily scale, but this is actually part of the program that
needs to be achieved. 

In the mean time, I think that several strong messages emerge from the above remarks, that are particularly topical after the 2008 crisis:  \footnote{There are 
obviously many other
aspects that we leave aside. 
The destabilising use of leverage is one of them, see \cite{FGT} for a recent interesting paper.} 

\begin{itemize} 

\item Even liquid markets are not really liquid, and therefore have no reason to be efficient. One should stop taking market prices at face value, 
specially in many OTC markets where `liquidity' is deceptive. Quants should quit the obsession of exact calibration on market prices, in particular when the models are absurdly remote from reality. 
One of the worst example I know is the use of local volatility models, that are by construction able to fit any volatility surface -- 
so calibration will always work, and this is unfortunately why the approach is so popular.  
But using this framework to price more exotic derivatives using ``plain vanilla'' instruments can lead to disaster, even if plain vanilla markets
were efficient -- because the underlying reality has nothing to do with a local volatility process. The situation is obviously even worse if markets are not efficient. Errors are
propagated and amplified in a non-linear way, and the price and hedge of illiquid instruments can be totally nonsensical. There are many examples in the quantitative finance 
literature of erroneous models that can be easily calibrated, and that are therefore used and abused by financial engineers. In my view, the use of such models contributes to propagate systemic
risk, specially as they become standard practice.

\item Collective effects mediated by imitation or contagion pervade markets and lead to instabilities.
Prosperity relies heavily
on trust, which is an immaterial common good that has no inertia and can dissipate overnight \cite{trust}. The mechanisms that foster or destroy trust are intrinsically (or even tautologically) collective. 
The most efficient mechanism for contagion is through the dynamics of 
the price and of the order flow, which is public, common information. Since it is impossible to be immediately sure that a silly trade is indeed silly, its impact on the price can 
trigger an instability, as was likely the case during the flash crash of May 6th 2010. Being influenced by the behaviour of others seems to be one of the most common human trait, 
that persists across history. We are always worried that others may be smarter, or may have more information than we do. 
This imitation propensity is well known to lead to dramatic effects (see e.g. \cite{Keynes,Granovetter,Galam,Brock,Curty,Quentin,Nadal} and refs. therein), and must be one of the ingredients leading to crises and crashes \cite{AnimalSpirits}. The importance of hysteresis, in that respect, cannot be overemphasized \cite{Quentin,Nadal,PW,Lamba}.

There are many other contagion mechanisms -- we just mentioned the use of similar 
pricing and risk models. More generally,  common strategies lead to common 
positions (see e.g. the quant crunch of August 2007 \cite{Lo}), and so does the widespread diffusion of similar `toxic' products (e.g. CDOs). Benchmarking performance to the average of a peer group
promotes copy-cat behaviour. The cross-liability network between financial institutions or between companies can also be instrumental in wreaking havoc 
(see e.g. \cite{Battiston,Kuhn,Douady}).

\item Another important idea is that agents in financial markets are strongly heterogeneous. Physical systems where individual elements are both heterogeneous and strongly interacting are well known to be inherently fragile to small perturbations. These systems generically evolve in an intermittent way, with a succession of rather stable epochs punctuated by rapid, unpredictable changes -- again, even when the exogenous drive is smooth and steady. Within this metaphor of markets, competition and complexity could be the essential cause of their endogenous instability.
 
The main problem with the current theories is that they are based on the idea that one can replace ensemble of heterogeneous and interacting agents by a unique representative one, in other words that the micro- and macro-behaviour should coincide \cite{Kirman}. Within this framework, crises are expected 
to require a major external shock, whereas in reality small local disturbances can trigger large systemic effects (the US sub-prime market represented in itself only a minor fraction of the global credit market but still stoked a global crisis).

\item Finally, there are a number of explicit destabilising feedback loops that regulators should investigate and abate. Some are a direct consequence of the faith in the efficiency of
markets, such as the ``mark-to-market'' accounting rule, which relies on the idea that market prices are fair and unbiased. Such a pro-cyclical practice applied on credit derivatives 
has contributed to impair the balance sheet of many financial institutions in 2008, and amplified the mayhem. In my opinion, again, the ``fair price'' idea does not make sense without at
least the notion of an intrinsic uncertainty  and a liquidity discount based on a pessimistic estimate of the impact cost during a fire-sale. 
Other feedback loops are created by the use of financial derivatives (see \cite{Marsili, HommesD}) and/or, as alluded to above, by 
quantitative models themselves -- a vivid example in the crash of
1987 that was a direct consequence of the unwarranted trust in Black-Scholes' perfect replication theory. 

There are also nasty feedback loops lurking in
the high frequency, micro-structure side. We 
have mentioned several times in this paper the spread $\to$ volatility $\to$ spread loop that is probably at the origin of most ``spontaneous'' liquidity crises (such as the one of May 6th, 2010, but also all the daily jumps that we discussed but that rarely make the news). 
It would be interesting to investigate mechanisms that help averting those. For example, dynamic make/take fees that depend on market conditions and on the distance between the placed order and the last traded price 
could endogeneize stabilising feedback loops. This is clearly an issue around which academic research and regulation merge, which makes modelling high frequency so exciting.
\end{itemize}

Whether or not the above ingredients can be mixed and tuned to provide a truly quantitative theory of economic and financial crises remains of course, at this 
stage, a fascinating open problem.

\vskip 0.5cm 
{\bf Acknowledgements:} I thank Arthur Berd for inviting me to assemble my thoughts and review my own work on these topics (this explains why the
reference list is shamefully self-serving). I also thank all my 
collaborators on these subjects, who help me 
shape my understanding of markets, in particular C. Biely, C. Borghesi, L. Borland, Z. Eisler, J. D. Farmer, J. Kockelkoren, Y. Lemperi\`ere, F. Lillo, M. Potters and M. Wyart.
I also enjoyed discussions, over the years, with X. Gabaix, J. Gatheral \& M. Marsili.

\end{document}